# IMPROVED SECURE ADDRESS RESOLUTION PROTOCOL

Abhishek Samvedi[1], Sparsh Owlak[2] and Dr. Vijay Kumar Chaurasia[3]

[1]Indian Institute of Information Technology, Allahabad, India
abhisheksmvd@gmail.com
[2]Indian Institute of Information Technology, Allahabad, India
maadhav.owlak@gmail.com
[3]Indian Institute of Information Technology, Allahabad, India
vijayk@iiita.ac.in

*ABSTRACT*

*In this paper, an improved secure address resolution protocol is presented where ARP spoofing attack is prevented. The proposed methodology is a centralised methodology for preventing ARP spoofing attack. In the proposed model there is a central server on a network or subnet which prevents ARP spoofing attack.*

## 1. INTRODUCTION

ARP spoofing has become a major problem in the present scenario. ARP spoofing can lead to many other attacks like man in the middle attack in secure socket layer. Thus steps must be taken to prevent this type of attack. In this paper a scheme is proposed to prevent ARP spoofing attack.

Section I gives a brief introduction of the situation. Section II discusses the proposed solution for preventing ARP spoofing attack in detail. Section III discusses the message formats used in proposed scheme. Section IV discusses the performance evaluation of proposed scheme against standard ARP protocol. Section V summarizes and concludes the paper.

### 1.1 Address Resolution Protocol

The Address Resolution Protocol (ARP) is used by Internet Protocol [IP], defined in [RFC826], to bind the IP addresses to the MAC addresses, which is stored in ARP cache of each client machine. This protocol works on network layer. In IPv6 this functionality is provided by Neighbour Discovery Protocol (NDP).

### 1.2 ARP Poisoning

ARP Poisoning is done by sending ARP Reply packet to victim node with sender's (attacker) IP address and MAC address as destination IP and MAC address respectively, as shown in figure 1. The victim when processes the ARP Reply packet, this brings change in its ARP table for destination IP address with attacker's MAC address, which causes the victim to send all packets destined to target host to attacker. The attacker then can read and modify packets flowing between target node and victim node.

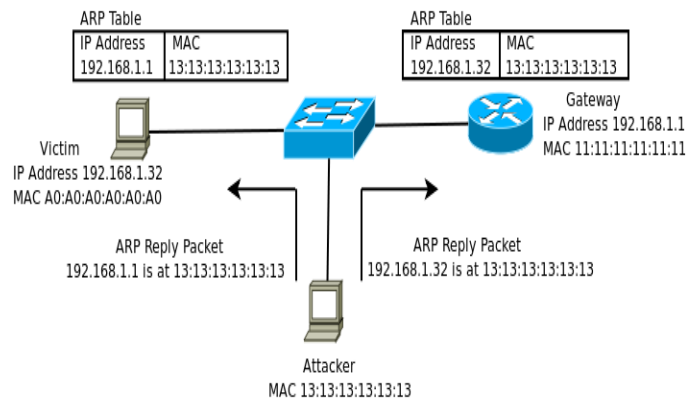

Figure 1. ARP poisoning

### 1.3 Dynamic Host Configuration Protocol

The Dynamic Host Configuration Protocol (DHCP) is a protocol, which provides IP address to client from IP address pool, which in return results into the connectivity of client with rest of the network. It involves clients and a server operating in a client-server model. When the DHCP server receives a request from a client, server determines the network to which the client is connected, and allocates an IP address or prefix and configuration information to the client. DHCP server grants IP addresses to clients for a limited period and clients are responsible for renewing their IP address before that time expires. DHCP is used in IPv4 and IPv6. This protocol is defined in [RFC2131].

## 2. RELATED WORK

Previous work relating ARP security are S-ARP[1], Ticket based ARP[2], Enhanced ARP[3], securing unicast protocol by extending DHCP[4] and a centralised detection and prevention technique against ARP poisoning[5]. In S-ARP each host has a public private key pair and a digital certificate obtained from a central certificate authority (CA) which is local to the organisation. Each host has to register its IP address, and public key (contained in certificate) to the authoritative key distributer (AKD), as every host has to get a certificate the total overhead incurred by the local CA will be high to verify every new host. In this scheme as every host has its certificate the chances of such attack are high where an attacker steals the certificate of a host and pretends to be that victim host on the network and performs malicious activities. In S-ARP for the case of dynamic networks the DHCP server has to first consult the AKD server before providing the new host an IP address. This means that a new host has to register an entry with AKD server first which has to be done manually as the new host does not have an IP address at this stage. Also the communication between new host and local CA will be done manually in dynamic network situation. In Ticket based Address Resolution Protocol [2] research paper there is a local Local Ticket Agent (LTA) which issues tickets to new host. Thereafter the communications in ARP protocol are done using the ticket concept. Here as every new host requires a ticket, overhead incurred will be high. Another possible attack here is that an attacker makes a copy of a ticket of genuine host and then uses that ticket for malicious activities on the network. In Enhanced ARP[3] research paper each host maintains a long term IP-MAC mapping table apart from ARP cache. This IP-MAC mapping table contains IP-MAC mapping of all hosts on its network. This scheme involves too much memory usage and thus cost incurred will also be high. Also the updating of IP-MAC mapping table for each host will be difficult. In dynamic networks the IP addresses provided by DHCP server to various hosts have to be renewed after some time thus frequent updating of IP-MAC table will be required which will require enough overhead in this scheme. In the research paper securing unicast protocol by extending DHCP [4], the DHCP man in the middle attack is possible. In the research paper a centralised detection and prevention technique against ARP poisoning [5] the ACS (ARP central server) acts very much similar to DHCP server and is vulnerable to DHCP man in the middle attack.

## 3. PROPOSED SOLUTION

The proposed scheme which removes ARP spoofing attack is for dynamic networks, where there are hosts, and a new entity called Central Server. The Central Server maintains an IP-mac table for the subnet or the network it is present in. This IP-mac table contains information of IP-mac binding information of all hosts on the subnet or network the central server is present in and which have been allocated IP address by the DHCP server nearest to this central server. Relay Agents can be used in case of large networks. Whenever a host on a subnet or network is allocated an IP address by the DHCP server, it also informs the central server on that subnet or network through IP_send message. This message is sent on data link layer. It is signed by the secret key shared between DHCP server and central server. The central server then sends a message IP_reply to the DHCP server to show acknowledgement of IP_send message. This message is sent on data link layer and is signed by the central server using the symmetric key shared between central server and DHCP server. In a network all the ARP request and ARP reply messages will be sent to this Central Server. The client or hosts will not communicate the ARP request and ARP reply messages to each other. The scenario of a network can be shown as:-

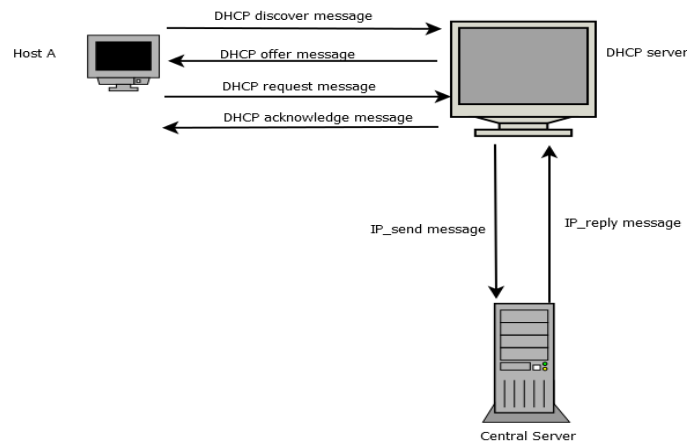

Figure 2. Proposed subnet or network setup

The procedure for entry of a new host in a subnet or network will be as follows:-
1. The new host broadcasts the DHCP discover message containing the MAC address of the host.
2. The DHCP server allocates an IP address to this new host following the standard DHCP protocol.
3. The DHCP server then sends an IP_send message to the central server. This message is sent on data link layer. It is digitally signed by the symmetric secret key shared between DHCP server and central server.
4. The central server after receiving this message will update its IP-mac table. The frame format of IP_send message is discussed in the following section.
5. After this the central server will send IP_reply message to the DHCP server showing acknowledgement of IP_send message. This message is sent on data link layer and is digitally signed by symmetric key shared between central server and DHCP server. The frame format of IP_reply message is discussed in the following section.

### 3.1 Prevention of ARP spoofing attack

In this scheme all the ARP reply and ARP request messages are send to Central Server. When the Central Server receives an ARP request message it provides the requesting host with the MAC address of corresponding IP address it wants by sending ARP reply message which is again digitally signed by the central server. In this case asymmetric cryptography is used. The digital certificate of central server is also attached in this ARP reply message. This digital certificate can be obtained from public certificate authority like VeriSign. One digital certificate is required and it can be distributed to all central servers. When the Central Server receives an ARP reply message which is sent by a host which wants to get its IP-mac combination information changed due to change in its MAC address then the Central Server checks it's IP-mac table and sends 50 ARP_Check messages to the previous MAC address stored as a combination with the IP address provided. These ARP_Check messages are again digitally signed by the Central Server. The Central Server also attaches its digital certificate with these messages. These ARP_Check messages are sent on data link layer and its frame format is discussed in the following section. If the Central Server gets a reply even to any one of these messages then it keeps the previous IP-mac combination in IP-mac table otherwise it changes the previous entry with new entry. In case the Central Server gets an ARP reply message from previous MAC address it sends an ARP_NoChange message to the MAC address which initiated this procedure. This ARP_NoChange is digitally signed by central server with digital certificate attached with this message. In case of Denial of Service (DOS) attack on the system with MAC address which has to be changed in the IP-mac table of Central Server, at least one ARP reply will be returned by it to the Central Server with a probability of 99.5% (=1-0.95$^{50}$). Thus sending 50 ARP_Check messages just increases the probability that we will get some reply from a host which has the previous MAC address which is to be changed in the case of ARP spoofing attack. On the other hand if the Central Server does not gets an ARP reply message from previous MAC address it sends an ARP_Ack message to the MAC address which initiated this procedure. This message indicates the client that appropriate modification in IP-MAC table in central server has been done. This message is again digitally signed by central server with digital certificate attached.

In this situation two types of ARP spoofing attack can be thought of:-
    **3.1.1** The attacker sends a fake ARP reply message to Central Server asking it to change its IP-mac combination in IP-mac table. In this situation the Central Server uses the checking scheme of sending 50 ARP_Check messages to previous MAC address as already discussed. This checking scheme prevents false entry in IP-mac table of Central Server thus preventing ARP spoofing attack.

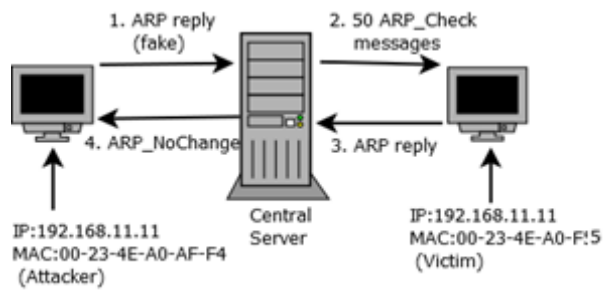

Figure 3: first possible type of ARP spoof attack

    **3.1.2** In this case a genuine host wants to get its IP-mac combination stored in IP-mac table in Central Server to get changed because of change in its MAC address, thus it will send the ARP reply message to Central Server for the same. Here in this case the Central Server will again follow its checking procedure. At this step the attacker knowing the previous MAC address of the genuine host who wants to get its IP-mac combination changed will falsely send an ARP reply message to the Central Server using the previous MAC address of genuine host. Thus the request of genuine host will get cancelled. Still In this situation the ARP spoofing attack is prevented because the Central Server will send a digitally signed ARP_NoChange message to the host which initiated this procedure. Thus this genuine host can again request a new IP address from Secure-DHCP server.

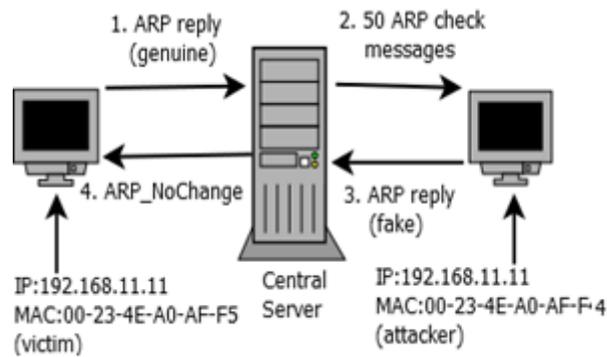

Figure 4. Second possible type of ARP spoof attack

Other possible attack in this scenario is the DOS attack on central server by a number of fake ARP reply messages. This DOS attack can be prevented by monitoring mechanism using Intrusion Detection System (IDS).

Also a situation arises that an attacker is performing a DOS attack on central server by sending a number of fake ARP reply messages to central server. Now at the same time a genuine user sends ARP reply message to central server then this message will also be discarded by the central server in order to mitigate the DOS attack. However since the client will not receive ARP_Ack message from central server it can conclude that its request has not been accepted so it can again send ARP_reply message to central server after some time.

## 4. MESSAGE FORMAT

The messages IP_send, IP_reply, can be send on data link layer with adequate changes in frame format of standard protocols used like Ethernet II to accommodate the required information. The messages IP_send, IP_reply, are messages communicated in a network amongst DHCP server, Central Server so these messages can be communicated on data link layer.  Relay agents can be used in case of large networks. The messages ARP_Check, ARP_NoChange, ARP_Ack can be also be send on data link layer. ARP request and ARP reply follow the standard ARP protocol frame format. To make the scheme compatible with networks following the normal standard protocols the gateways should be modified with extra capability to identify traffic flowing from outside network or inside network so that the gateway can follow the required network protocols to forward the traffic. The possible frame format for these messages is as follows:

### 4.1 IP_send message

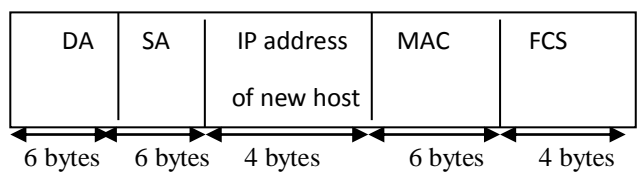

Figure 5. Frame Format of IP_send message

The fields in this frame format are:
- **4.1.1** Destination Address (6 bytes): This is the MAC address of destination system (CentralServer).
- **4.1.2** Source Address (6 bytes): this is the MAC address of source system (Secure DHCP).
- **4.1.3** IP address (4 bytes): This is the IPv4 address of the new host.
- **4.1.4** MAC address of new host (6 bytes): This is the MAC address of the new host.
- **4.1.5** FCS (4 bytes): This field stands for frame check sequence.

## 4.2 IP_reply message

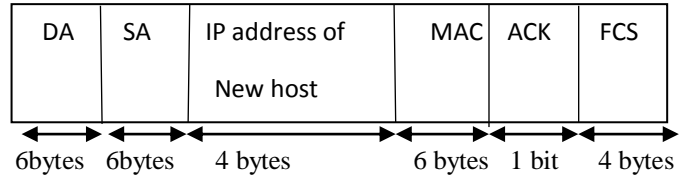

Figure 6. Frame Format of IP_reply messages

The fields in this frame format are:
- **4.2.1** Destination Address (6 bytes): This is the MAC address of destination system (SecureDHCP).
- **4.2.2** Source Address (6 bytes): This is the MAC address of source system (CentralServer).
- **4.2.3** IP address (4 bytes): This is the IPv4 address of the new host.
- **4.2.4** MAC address (6 bytes): This is the MAC address of new host.
- **4.2.5** ACK (1 bit): This is a 1 bit field showing the acknowledgement of IP address by new host. This bit is set to 1 to show acknowledgement of IP address.
- **4.2.6** FCS (4 bytes): This field stands for frame check sequence.

## 4.3 ARP_CHECK message

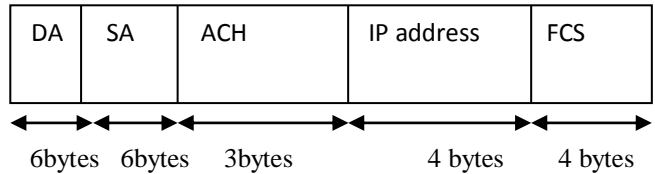

Figure 7. Frame Format of ARP_CHECK messages

The fields in this frame format are:
- **4.3.1** Destination Address (6 bytes): This is the MAC address of destination system (The host (if exists) with the MAC address stored in IP-mac table which has to be replaced with new MAC address).
- **4.3.2** Source Address (6 bytes): This is the MAC address of source system (CentralServer).
- **4.3.3** ACH field (3 bytes): This field will store the string ACH which indicates ARP_CHECK message.
- **4.3.4** IP address (4 bytes): This is the IPv4 address of the host whose MAC address has to be changed.
- **4.3.5** FCS (4 bytes): This field stands for frame check sequence.

## 4.4 ARP_NoChange message

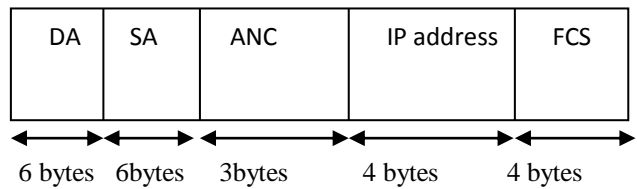

Figure 8. Frame Format of ARP_NoChange

The fields in this frame format are:
- **4.4.1** Destination Address(6 bytes):This is the MAC address of destination system ( The host which initiated the ARP reply message to Central Server with the need to make changes in the IP-mac table and update it with new MAC address of this host ).
- **4.4.2** Source Address (6 bytes): This is the MAC address of source system (CentralServer).
- **4.4.3** ANC field (3 bytes): This field will store the string ANC which indicatesARP_NoChangemessage.
- **4.4.4** IP address (4 bytes): This is the IPv4 address of the host which initiated the request to update the IP-mac table.
- **4.4.5** FCS (4 bytes): This field stands for frame check sequence.

### 4.5 ARP_Ack message

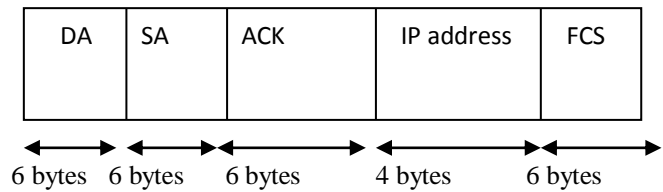

Fig 9. Frame Format of ARP_Ack

The fields in this frame format are:
- **4.5.1** Destination Address(6 bytes): This is the MAC address of destination system (The host which Initiated the ARP reply message to Central Server with the need to make changes in the IP-mac table and update it with new MAC address of this host).
- **4.5.2** Source Address (6 bytes): This is the MAC address of source system (CentralServer).
- **4.5.3** ACK field (3 bytes): This field will store the string ACK which indicates ARP_Ackmessage.
- **4.5.4** IP address (4 bytes): This is the IPv4 address of the host which initiated the request to update the IP-mac table.
- **4.5.5** FCS (4 bytes): This field stands for frame check sequence

## 5. PERFORMANCE EVALUATION

We will evaluate the performance of proposed scheme against standard ARP protocol and standard DHCP protocol.

### 5.1 Performance Evaluation of the proposed scheme against DHCP protocol

The figure below shows the scenario where IP address is provided to new host using standard DHCP protocol. This scenario involves just one DHCP server.

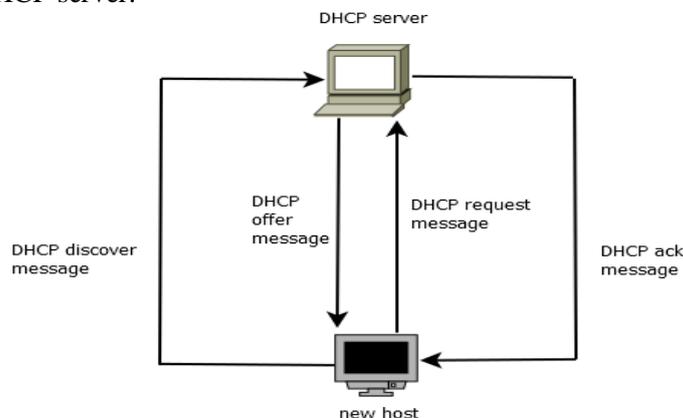

Figure 10. Performance evaluation of DHCP protocol

As is seen from Figure 10 the total transaction of messages involved in this process is 4.

Now we evaluate our proposed scheme. The figure below shows the situation:

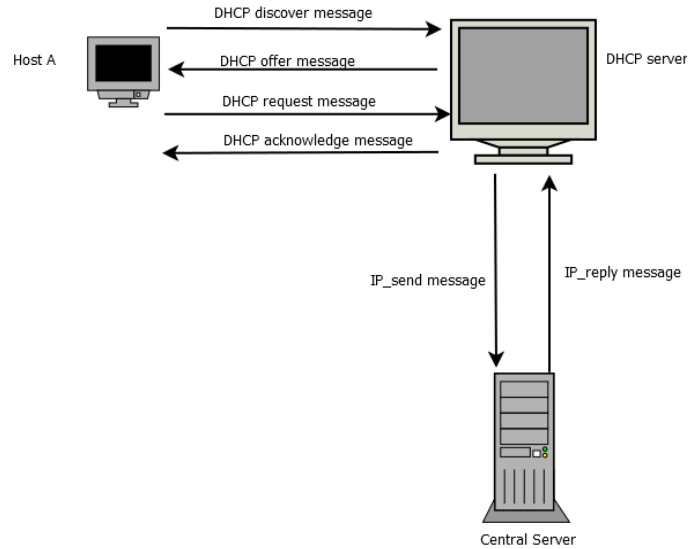

Figure 11. Performance evaluation of proposed scheme

As is seen from figure 11 that total transaction of messages involved is 6,
Thus if we compare the performance of standard DHCP protocol and the proposed scheme, we find that the standard DHCP protocol is slightly better than our proposed scheme in terms of cost involved in transaction of messages but our proposed scheme also removes the ARP spoofing attack thus making the network secure.

### 5.2 Performance Evaluation of proposed scheme against ARP protocol

To evaluate performance of proposed scheme against standard ARP protocol we consider two situations:

**5.2.1** The host A wants to know MAC address of host B: In the case of standard ARP protocol the host A will broadcast ARP request message on the network and will wait for ARP reply from host B. The situation is shown in figure below:

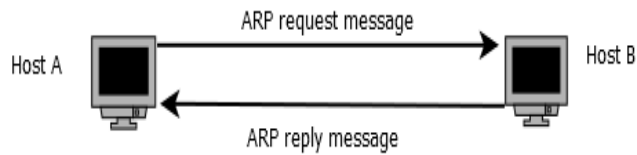

Figure 12. Performance evaluation of ARP protocol

As is seen from figure 12 the total transaction of messages involved in this case is 2.
If we consider the same situation in a subnet or network following our proposed scheme then the host A will send the ARP request message to the Central Server. The Central Server will reply with ARP reply message to host A providing MAC address of host B. The situation is shown below:

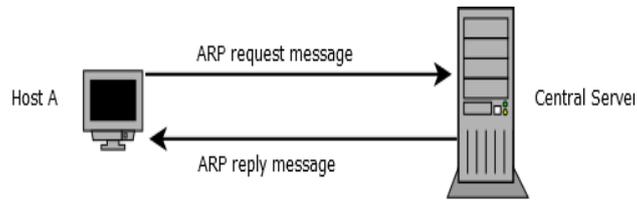

Figure 13. Performance evaluation of proposed scheme

As is seen from figure 13 the total transaction of messages involved is 2.
Thus in this situation we see that performance of proposed scheme is equivalent to standard ARP protocol in terms of cost involved in transaction of messages.

**5.2.2** Host A has got its MAC address changed and it wants to inform other hosts on the network: In this situation if the network follows standard ARP protocol then host A broadcasts ARP reply message to all other hosts on the network containing the information of its changed MAC address. The situation is shown in figure below.

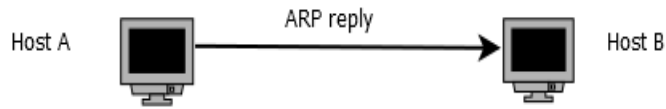

Fig 14. Performance evaluation of ARP protocol

In this situation in the case of standard ARP protocol the total transaction of messages is just 1.
If such a situation occurs in a network following the proposed scheme then the situation can be shown by the figure:

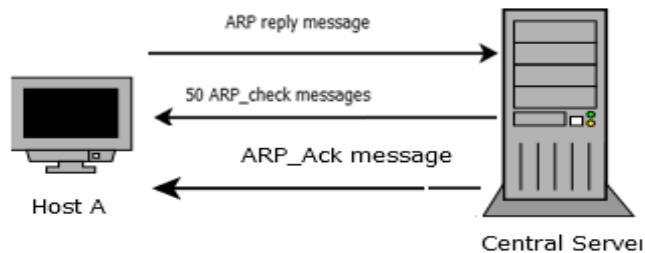

Figure 15. Performance evaluation of proposed scheme

In this situation we see that the host A sents ARP reply message to Central Server for updation of IP-mac table. In response to this the Central Server first sends 50 ARP_check messages to the previous MAC address. If it gets no reply to any of these messages then it updates the IP-mac table and finally the ARP_Ack message is sent from central server to the client as an acknowledgement for the change.
Total Transaction of messages involved in this situation is 52 messages.
Here we observe that the total cost involved in transaction of messages in the case of our proposed scheme is more than the standard ARP protocol , however the proposed scheme makes a network secure and ARP spoofing attack is not possible in it.

## 6. CONCLUSION

The proposed scheme removes ARP spoofing attack in a dynamic network. The scheme is highly secure and good for dynamic networks. The scheme is also compatible with existent networks following standard network protocols with some modifications in gateways. DHCP denial of service attack can also be removed if we follow the monitoring mechanism which is generally followed to prevent such attacks. Thus the proposed scheme is highly secure.

## 6. REFERENCES

[1] D. Bruschi, A. Ornaghi, and E. Rosti, "S-arp: a secure address resolution protocol," in Computer Security Applications Conference, 2003. Proceedings. 19th Annual. IEEE, 2003, pp. 66–74.

[2] W. Lootah, W. Enck, and P. McDaniel, "Tarp: Ticket-based address resolution protocol," vol. 51, no. 15. Elsevier, 2007, pp. 4322–4337.

[3] S. Nam, D. Kim, and J. Kim, "Enhanced arp: preventing arp poisoning based man-in-the-middle attacks," Communications Letters, IEEE, vol. 14, no. 2, pp. 187–189, 2010.

[4] B. Issac and L. Mohammed, "Secure unicast address resolution protocol (s-uarp) by extending dhcp," in Networks, 2005. Jointly held with the 2005 IEEE 7th Malaysia International Conference on Communication. 2005 13th IEEE International Conference on, vol. 1. IEEE, 2005, pp. 6–pp.

[5]Sumit Kumar and ShashikalaTapaswi, "A centralized detection and prevention technique against ARP poisoning.",IEEE,pp. 259-264.

[6]Internet Engineering Task Force. (2004) Dynamic Host Configuration Protocol [Online]. Available from: http://tools.ietf.org/html/rfc2131 [Accessed 24rd September,2013].

**Authors**

1. Abhishek Samvedi is a student of Master of Science. His domain of M.S. program is CLIS (Cyber law and information security) from IIIT-Allahabad.

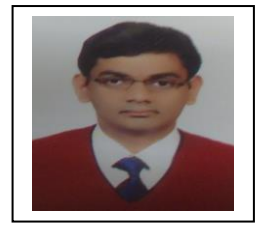

2. Sparsh Owlak is a student of Master of Science. His domain of M.S. program is CLIS (Cyber law and information security) from IIIT-Allahabad.

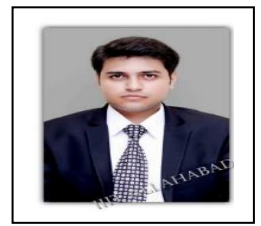

3. Vijay Kumar Chaurasiya is Doctor of philosophy from IIIT-Allahabad. His area of Specialization Is in Wireless and mobile network.

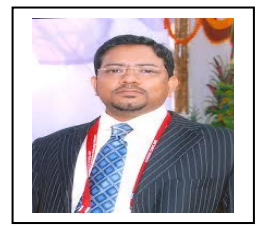